\documentclass[aps,prb,twocolumn,unsortedaddress,floatfix,nofootinbib,superscriptaddress]{revtex4-2}
\usepackage{amsthm}
\usepackage{amsfonts}
\usepackage{siunitx}
\usepackage{amsmath}
\usepackage{amssymb}
\usepackage{graphicx}
\usepackage{verbatim}
\usepackage[colorlinks]{hyperref}
\usepackage{tikz}
\usepackage{braket}
\usepackage{xcolor}

\makeatletter
\def\@ssect@ltx#1#2#3#4#5#6[#7]#8{%
  \def\H@svsec{\phantomsection}%
  \@tempskipa #5\relax
  \@ifdim{\@tempskipa>\z@}{%
    \begingroup
      \interlinepenalty \@M
      #6{%
       \@ifundefined{@hangfroms@#1}{\@hang@froms}{\csname @hangfroms@#1\endcsname}%
       {\hskip#3\relax\H@svsec}{#8}%
      }%
      \@@par
    \endgroup
    \@ifundefined{#1smark}{\@gobble}{\csname #1smark\endcsname}{#7}%
  }{%
    \def\@svsechd{%
      #6{%
       \@ifundefined{@runin@tos@#1}{\@runin@tos}{\csname @runin@tos@#1\endcsname}%
       {\hskip#3\relax\H@svsec}{#8}%
      }%
      \@ifundefined{#1smark}{\@gobble}{\csname #1smark\endcsname}{#7}%
      \addcontentsline{toc}{#1}{\protect\numberline{}#8}%
    }%
  }%
  \@xsect{#5}%
}%
\makeatother

\definecolor{linkcolor}{RGB}{0,83,166}
\hypersetup{
  colorlinks = true,
  allcolors = {linkcolor}
}

\newcommand{\chiqmps}{\chi^\text{MPS}_\text{Q}}

\newcommand{\chigtmps}{\chi^\text{MPS}_\text{GT}}

\newcommand{\lmax}{l_\text{max}}

\newcommand{\epsmpstns}{\epsilon_\text{MPS}^\text{TNS}}
\newcommand{\epsmpsqa}{\epsilon_\text{MPS}^\text{QA}}
\newcommand{\epsqatns}{\epsilon_\text{QA}^\text{TNS}}

\begin{document}
\newcommand{\mytitle}{Evaluating classical simulations with a quantum processor}
\title{\mytitle}

\newcommand{\affildw}{D-Wave Quantum Inc., Burnaby, British Columbia, Canada}

\newcommand{\affilsfu}{Department of Physics, Simon Fraser University, Burnaby, British Columbia, Canada}
\newcommand{\affilubc}{Department of Physics and Astronomy and Quantum Matter Institute, University of British Columbia, Vancouver, British Columbia, Canada}

\author{Alberto Nocera}
\email[]{alberto.nocera@ubc.ca}
\affiliation{\affilubc}
\author{Jack Raymond}

\affiliation{\affildw}

\author{William Bernoudy}

\affiliation{\affildw}

\author{Mohammad H.~Amin}
\affiliation{\affildw}
\affiliation{\affilsfu}

\author{Andrew D.~King}
\email[]{aking@dwavesys.com}
\affiliation{\affildw}

\date{\today}
\begin{abstract}
  As simulations of quantum systems cross the limits of classical computability, both quantum and classical approaches become hard to verify.  Scaling predictions are therefore based on local structure and asymptotic assumptions, typically with classical methods being used to evaluate quantum simulators where possible.  Here, in contrast, we use a quantum annealing processor to produce a ground truth for evaluating classical tensor-network methods whose scaling has not yet been firmly established.  Our observations run contrary to previous scaling predictions, demonstrating the need for caution when extrapolating the accuracy of classical simulations of quantum dynamics.  Our results demonstrate that the virtuous cycle of competition between classical and quantum simulations can lend insight in both directions.
\end{abstract}

\maketitle

\def\title#1{\gdef\@title{#1}\gdef\THETITLE{#1}}

The recent advent of programmable many-body quantum systems has brought with it a competition between quantum and classical methods seeking to delineate the boundary between classical computability and quantum advantage~\cite{preskill_quantum_2012}. 
This is especially true in quantum simulation~\cite{daley_practical_2022}, where available quantum processing units (QPUs) can play to their strengths~\cite{morvan_phase_2024,king_beyondclassical_2025,andersen_thermalization_2025,haghshenas_digital_2025} and where novel classical techniques based on tensor networks~\cite{berezutskii_tensor_2025} have found growing success~\cite{tindall_confinement_2024,begusic_fast_2024,tindall_dynamics_2025,park_simulating_2025,rudolph_simulating_2025}.
Any claim of quantum simulation beyond the reach of classical methods, for example, requires as a basis step that small simulations be validated classically.  Once a low error level has been established in small quantum simulations, one must make the case indirectly that the quantum simulation is accurate in large systems.  Such indirect techniques include, for example, the use of Clifford gates in random circuit sampling~\cite{boixo_characterizing_2018} or Kibble-Zurek scaling in simulation of critical dynamics~\cite{king_beyondclassical_2025}.  

Among classical algorithms, methods to optimize and time evolve matrix product states (MPS) are notable for their controlled approximation error, which can be systematically reduced to an arbitrarily low level by increasing a single parameter, the bond dimension, albeit at increased computational cost and memory requirements. Scaling characteristics of these methods are also well understood through the area law of entanglement, which explains why MPS efficiently captures low-entanglement states common in one-dimensional systems. Because of this controllability and theoretical grounding, MPS methods are particularly well-suited for obtaining ground-truth in benchmarking and validation studies. This level of control is not generally available in other algorithms such as projected entangled pair states (PEPS) or neural quantum states (NQS), especially when uncontrolled approximations are introduced to reduce the cost of tensor network contraction or optimization, for example.  Such approximations often obscure the reliability of the results.

When simulating locally coupled Hamiltonian dynamics, a fixed length of evolution time (a fixed circuit depth in the discretized case) imposes a limit on the range of correlations. This in turn provides insights into the asymptotic scaling behavior of some classical simulation algorithms. However, predicting the system size at which this asymptotic scaling sets in is nontrivial, as it depends sensitively on the growth of quantum correlations. In particular, the range of quantum correlations---often linked to out-of-time-order correlators (OTOCs)---can differ significantly from spin-spin correlation length. For example, in random circuit sampling, spin-spin correlation length is almost zero, but OTOCs can be system spanning. 
It is therefore important to develop methods for assessing the accuracy of classical algorithms and validating their scaling assumptions in regimes where MPS ground-truth solutions are no longer accessible.

In this paper, we aim to reverse the conventional approach by using a quantum processor to evaluate the performance of a classical simulation algorithm, rather than using classical algorithms to benchmark quantum hardware. Specifically, we demonstrate how a quantum annealing (QA) processor can be employed to assess the accuracy of novel tensor-network methods. This is possible because the accuracy of quantum processors and the moderate scaling of their error with system size and annealing time (quench duration) is now well established~\cite{king_beyondclassical_2025}. As a result, quantum annealing processors can serve as a reliable source of ground-truth data for benchmarking classical algorithms.  By leveraging the statistical independence between classical and quantum errors, we further extend our analysis into regimes where these errors become comparable in magnitude.  Our findings contradict a recent prediction about error scaling in a novel classical algorithm based on belief-propagation-gauged tensor-network states (BP-TNS)~\cite{tindall_dynamics_2025}, both in regimes where BP-TNS and QA errors are comparable in a quantum dynamics simulation~\cite{king_beyondclassical_2025}, and in regimes where QA error remains significantly lower than BP-TNS.

\section{Tensor Networks and Entanglement}

In simulations of quantum dynamics, entanglement plays a central role~\cite{amico_entanglement_2008}, embodying the leading computational bottleneck for classical tensor networks.  While generic Hilbert space states are highly entangled yet physically uninformative, physically relevant states---such as ground states and low-energy excitations of local Hamiltonians---reside in a structured, low-entanglement corner of the Hilbert space that can be approached with efficient classical approximations~\cite{poulin_quantum_2011,vidal_efficient_2003}.

More precisely, these states are characterized by the so-called area law, where the entanglement entropy of a block of qubits scales with the size of its boundary rather than with its volume, due to the locality of Hamiltonian interactions. This property underpins the success of MPS and density-matrix renormalization group (DMRG) algorithms for ground-state calculations in one spatial dimension~\cite{schollwock_densitymatrix_2011}. In higher dimensions, MPS methods are inefficient even for area-law entangled states.  However, higher-dimensional tensor network analogues of MPS---PEPS, also known as tensor network states (TNS)\footnote{The terms TNS and PEPS are used interchangeably in the community, so we follow the methods' original naming.}---can often provide efficient approximations for representing area-law states.  This is not always the case~\cite{ge_area_2016}, and when these representations exist, they cannot always be contracted efficiently, which is required, for example, when computing norms and measuring local observables. Indeed, because of the presence of loops, \emph{exact} PEPS contraction is \#P-complete~\cite{PhysRevResearch.2.013010}, and several \emph{approximate} algorithms have been proposed to overcome this problem, both for ground state and time dynamics~\cite{Haghshenas2019,Evenbly2018,tindall_gauging_2023,jiang_accurate_2008,Murg2007,Orus2009,Xie2009,dziarmaga_time_2021}.

\begin{figure*}
  \includegraphics[scale=.8]{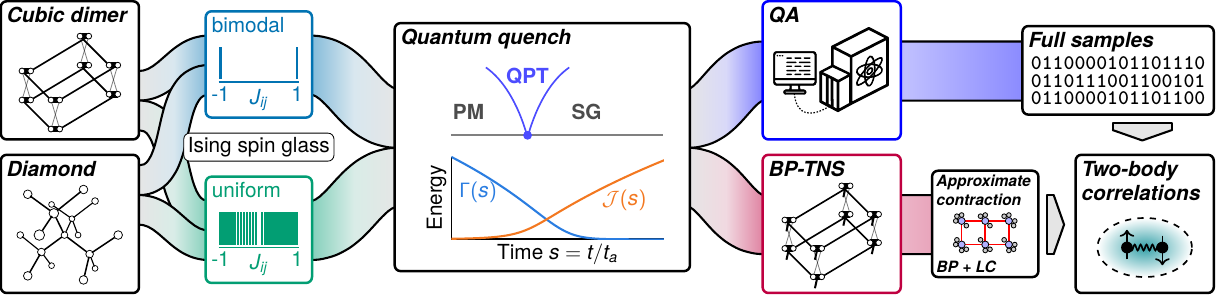}
  \caption{{\bf Quantum dynamics of glassy 3D Ising models.}  Random spin-glass realizations are generated by sampling couplings from bimodal or uniform (discretized, nonzero) distributions.  Systems are quenched through the QPT separating paramagnet and spin-glass phases.  Time evolution is performed using real quantum dynamics (QA) or Trotterized dynamics (BP-TNS).  Two-body correlations are estimated using Monte Carlo sampling (QA) or tensor-network contraction (BP-TNS).}\label{fig:1}
\end{figure*}

A recent study~\cite{king_beyondclassical_2025} demonstrated that quantum states following a finite-duration quench of a two- or three-dimensional Ising spin glass in a QA processor exhibit area-law entanglement scaling.  This was established by comparing QA data against MPS methods (MPS time-evolved with the time-dependent variational principle (TDVP)~\cite{haegeman_timedependent_2011,haegeman_unifying_2016}, i.e., MPS-TDVP).  In the same work, PEPS simulations with neighborhood tensor updates (PEPS-NTU)~\cite{dziarmaga_time_2021} were used to simulate the quench dynamics.  For increasing quench duration, PEPS showed increasing error, with saturation in bond dimension, despite decreasing entanglement in the system; this is attributed to the growing length scale of quantum correlations.

In Ref.~\cite{tindall_dynamics_2025}, a new class of BP-TNS methods was used to simulate the same experiment. In two spatial dimensions with cylindrical boundary conditions, contractions facilitated by a MPS message-passing BP algorithm showed a clear improvement over the PEPS-NTU method used in~Ref.~\cite{king_beyondclassical_2025} for the estimation of two-body correlations. In three spatial dimensions, or even in a torus, a different approximate contraction scheme was required to measure operators from the final time-evolved wave function. This was done using BP with loop corrections~\cite{alkabetz_tensor_2021,tindall_gauging_2023,evenbly_loop_2024} in cubic dimer and diamond lattices with $N\leq 54$ sites, displaying in some cases lower error than QA~\cite{tindall_dynamics_2025}.  For the biclique problems examined in Ref.~\cite{king_beyondclassical_2025}, such loop corrections are ineffective because the graph structure leads to a proliferation of loops. 

Even at this small scale of around 50 sites, it can be challenging to produce a precise and well-controlled ground truth using MPS, due to the stretched-exponential scaling of resource requirements in area-law states~\cite{king_beyondclassical_2025}.  It is therefore difficult to establish the onset of asymptotic scaling in system size expected from BP-TNS at fixed quench duration~\cite{tindall_dynamics_2025,rudolph_simulating_2025}, which may require analysis of instances that are too large to approach with MPS.  Here we approach this challenge using a QPU to generate an approximate ground truth.

\section{Simulation Task}

We consider the dynamics of the transverse-field Ising model (TFIM) whose time-dependent Hamiltonian interpolates between a driving Hamiltonian $\mathcal H_D$ and a classical Ising Hamiltonian $\mathcal H_P$ over a quench of duration $t_a$:
\begin{align}
  &\mathcal H(t/t_a) = \Gamma(t/t_a)\mathcal H_D + \mathcal J(t/t_a)\mathcal H_P,\\
&\mathcal H_D = \sum_{i=1}^n\sigma_i^x,\ \ \ \ \ \mathcal H_P = \hspace{-1ex}\sum_{1\leq i<j\leq n}\hspace{-1ex}J_{ij}\sigma_i^z\sigma_j^z.
\end{align}
In Ref.~\cite{king_beyondclassical_2025}, spin-glass couplings $J_{ij}$ were drawn from either a bimodal or uniform distribution (Fig.~\ref{fig:1}) following various geometries: square, diamond, cubic, and biclique---the latter two of which consist of two-qubit strongly-coupled dimers which are interconnected with random spin-glass couplings.  The bimodal cubic dimer case was previously used to demonstrate large-scale quantum-critical spin-glass dynamics in QA~\cite{king_quantum_2023} and the bimodal square case was recently used to investigate gap scaling using Monte Carlo methods~\cite{bernaschi_quantum_2024}.  The model annealing schedule $(\Gamma,\mathcal J)$ (Fig.~\ref{fig:1}) has, as its key feature in these disordered models, a transition from the transverse-field-aligned paramagnetic (PM) initial state through a quantum phase transition (QPT) into a spin-glass (SG) phase.  In Ref.~\cite{king_beyondclassical_2025} the accuracy of state sampling following a quench of duration $t_a$ was evaluated with respect to a ground truth generated with MPS-TDVP with high bond dimension $\chigtmps$.  Error was quantified based on $10^6$ samples using both Bhattacharyya distance and a normalized $\ell^2$ norm on two-body correlation errors
\begin{equation}
\epsilon_c = \sqrt{\frac{\sum_{i<j}(c_{ij}-\tilde c_{ij})^2}{\sum_{i<j}\tilde c_{ij}^2 }},
\end{equation}
with the tilde ($\tilde{\ }$) indicating ground-truth values.  Due to exponential sampling requirements for estimating the Bhattacharyya distance, $\epsilon_c$ was the preferred measure for distributional error over sampled states.  When tasked with matching QPU quality, MPS was found to require QPU-equivalent bond dimension $\chiqmps$ following a stretched-exponential scaling in system size, rendering it impractical for large simulations.  MPS, unlike BP-TNS, has well-controlled errors that tend to zero in the limit $\chi \rightarrow \infty$.

\begin{figure*}
\includegraphics{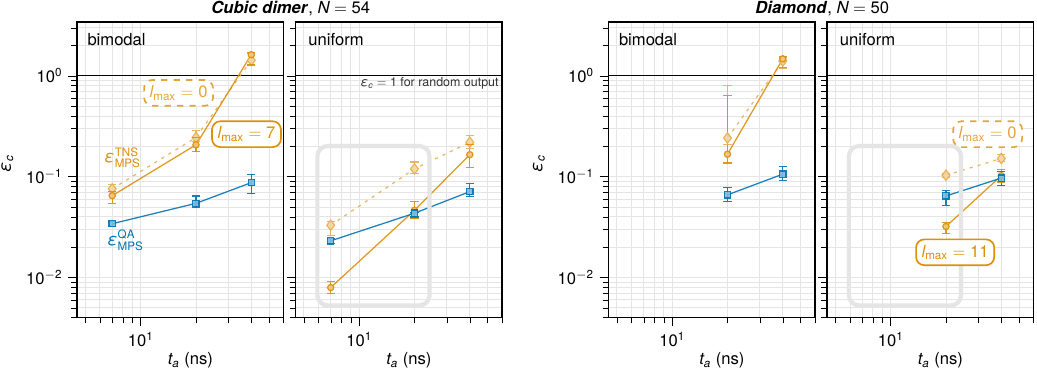}
\caption{{\bf Scaling of error with simulation depth.}  As $t_a$ increases from $\SI{7}{ns}$ to $\SI{40}{ns}$, both QA and BP-TNS ($\chi=8$ for cubic dimer, $\chi=16$ for diamond) show increasing median error in $20$-instance spin-glass ensembles.  Loop corrections are less beneficial for bimodal inputs and for longer $t_a$, increasing error in the most extreme cases.  Grey boxes indicate regime reported in Ref.~\cite{tindall_dynamics_2025}.  Error bars indicate 95\% bootstrap confidence interval on the ensemble median error. We have no MPS ground truth for $N=50$ diamond inputs at $t_a=\SI{7}{ns}$.}\label{fig:2}
\end{figure*}

This study was followed by improvements to BP-TNS methods: loop-correction expansions were proposed as a means of improving approximate tensor-network contraction~\cite{evenbly_loop_2024}, and this approach was applied directly in Ref.~\cite{tindall_dynamics_2025} to the TFIM simulation of Ref.~\cite{king_beyondclassical_2025}, in particular to uniform ensembles in diamond and cubic dimer geometries with $N \leq 54$.  In the large-system limit, a fixed $t_a$ (i.e., fixed depth of Trotterized circuit~\cite{rudolph_simulating_2025}) leads to a fixed correlation length $\xi$, which in turn suggests that $\epsilon_c$ should asymptote as a function of system size, with computational costs scaling only linearly.  This is in contrast to MPS, which scales poorly in system size but favorably in $\xi$.  However, the onset of the large-system limit, specifically concerning BP-TNS error, has not been established; this is the question we approach here.

\section{Scaling of errors in BP-TNS}

The BP-TNS algorithm is described in Ref.~\cite{tindall_dynamics_2025}, and consists of two main parts: time evolution of a TNS using BP-based simple update, and measurement of operators (in this case two-body correlations) via approximate contraction of the time-evolved TNS using loop-corrected BP.  Both here and in Ref.~\cite{tindall_dynamics_2025}, both steps are done using the same bond dimension $\chi$.

Locally treelike methods like BP-TNS (with and without loop corrections) are most effective at smaller energy scale and $t_a$, and quality is also sensitive to the coupling distribution.  In the case of uniform coupling distributions, many loops contain weak couplings that contribute proportionately small corrections, for which loop corrections are effective.  They are less effective in bimodal distributions, as already shown in geometrically-frustrated fullerenes~\cite{lopez-bezanilla_quantum_2025}.  Fig.~\ref{fig:2} shows ensemble-median $\epsilon_c$ with respect to the MPS ground-truth for BP-TNS ($\chi=8$ for cubic dimer, $\chi=16$ for diamond)\footnote{This matches results reported in Ref.~\cite{tindall_dynamics_2025}, mistakenly indicated as $\chi=6$ for cubic dimer inputs in early preprint versions~\cite{tindall_pc}.} and QA, denoted $\epsmpstns$ and $\epsmpsqa$ respectively, for cubic systems with $N=54$ sites ($3{\times}3{\times}3$ dimers, $L=3$) and diamond systems with $N=50$ sites ($5{\times}5{\times}8$ per notation in Ref.~\cite{king_beyondclassical_2025}).  QA correlations are estimated using up to $10^6$ samples, while BP-TNS and ground-truth (MPS) correlations are estimated via contraction of the operator expectation values.  BP-TNS results are shown for loop corrections on logical cubic subgraphs with up to seven bonds ($\lmax=7$) and diamond subgraphs with up to 11 bonds ($\lmax=11$), and without loop corrections ($\lmax=0$).  The quality of BP-TNS correlation estimates decays sharply relative to QA as a function of $t_a$.  For the weakest correlations (i.e., shortest $t_a$ in the uniform ensemble) BP-TNS can outperform QA.  For the strongest correlations (i.e., longest $t_a$ in the uniform ensemble) BP-TNS correlation estimates are worse than random sampling.  Moreover, the relative benefit of loop corrections decays in the same manner.

We would also like to track the scaling of error as a function of system size.  In particular, it was explicitly predicted~\cite{tindall_dynamics_2025} that error in cubic dimer systems should be higher for $L=3$ than for $L=4$ due to contributions around 3-loops spanning the periodic $z$-dimension that do not exist for $L=4$.  This prediction comes despite the observation of increasing $\epsmpstns$ up to $N=54$, so the trend for larger systems has not been clearly established.  This presents a challenge, because we cannot easily produce a ground truth for larger systems using MPS:  At $t_a=\SI{20}{ns}$ and $L=4$, $\chiqmps$ is projected to be around $3000$~\cite{king_beyondclassical_2025}; $\chigtmps$ should be significantly larger, leading to onerous memory requirements, for example exceeding the storage of a single node in the Frontier supercomputer.

Since a high-precision MPS ground truth appears impractical to obtain for $L=4$ at $t_a=\SI{20}{ns}$, we take another approach.  Namely, we use QA as a ground truth and consider $\epsqatns$, the correlation error of BP-TNS relative to a QA ground truth.  When QA is much more accurate than BP-TNS, this approach naturally gives a good estimate of BP-TNS error---and the converse is true when BP-TNS is much more accurate than QA.  When the two methods have comparable error, applying a QA ground truth is nontrivial; $\epsqatns$ may differ significantly from $\epsmpstns$.  But these error measures are simply renormalized $\ell^2$ norms, so we exploit two simplifying assumptions to estimate the error of BP-TNS simulations with respect to MPS ground truths when these are not available.

The first assumption, which is easily confirmed, is that QA and BP-TNS two-body correlation errors are for practical purposes statistically independent.  For example, the correlation coefficient for two-body correlation errors $c_{ij}-\tilde c_{ij}$ between QA and BP-TNS is $0.07$ (median instance) for uniform spin glasses at $t_a=\SI{20}{ns}$.  We therefore expect
\begin{equation}
\epsqatns \approx \sqrt{ \left(\epsmpsqa\right)^2 + \left(\epsmpstns\right)^2}.\label{eq:triangle}
\end{equation}

\begin{figure}
  \includegraphics{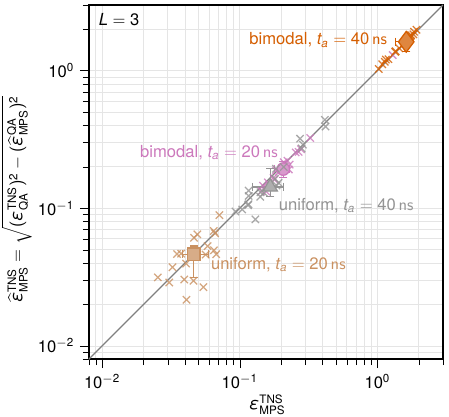}
  \caption{{\bf Direct and indirect measures of BP-TNS error in cubic dimer inputs.}  BP-TNS error $\epsmpstns$ ($\chi=8$, $\lmax=7$) is measured directly (horizontal axis) and estimated indirectly using Eq.~(\ref{eq:hat}) (vertical axis) for $L=3$ ensembles ($N=54$), with individual realizations indicated by ($\times$) and ensemble medians indicated by filled markers; error bars indicate 95\% bootstrap confidence interval.  Deviation is negligible when $\epsmpstns \gg \epsmpsqa$.  When $\epsmpstns \approx \epsmpsqa$, per-realization variance is significant but the ensemble median is in good agreement.}\label{fig:3}
\end{figure}

The second assumption is that $\epsmpsqa$ is approximately constant as a function of system size in QA.  Since QA follows physical dynamics with no confounding algorithmic factors such as loop corrections, this should be expected for large systems, and it was demonstrated for square lattices up to 90 sites in \cite{king_beyondclassical_2025} (Fig.~S25).  Accordingly, for a given ensemble and $t_a$, we define $\hat{\epsilon}{}_\text{MPS}^\text{QA}$ as $\epsmpsqa$ when we can compute it, and otherwise as the median value of $\epsmpsqa$ for the largest available size.  We use the largest available size ($L=3$ in this case) since finite-size effects are minimal and self-averaging is greatest.  Finally, we define the estimator for BP-TNS error:
\begin{equation}
\hat{\epsilon}{}_\text{MPS}^\text{TNS} =\sqrt{ \left(\epsqatns\right)^2 - \left( \hat{\epsilon}{}_\text{MPS}^\text{QA} \right)^2}.\label{eq:hat}
\end{equation}
This is well defined provided that QA and BP-TNS errors are sufficiently independent and QA is sufficiently accurate compared to BP-TNS.

In Fig.~\ref{fig:3}, $\hat{\epsilon}{}_\text{MPS}^\text{TNS}$ is shown to accurately estimate ensemble-median $\epsmpstns$ despite significant variation from one spin-glass realization to the next.  Naturally, in cases where $\epsmpstns$ is much larger than $\epsmpsqa$, for example in bimodal inputs, the estimator is very reliable. 

\begin{figure}
\includegraphics{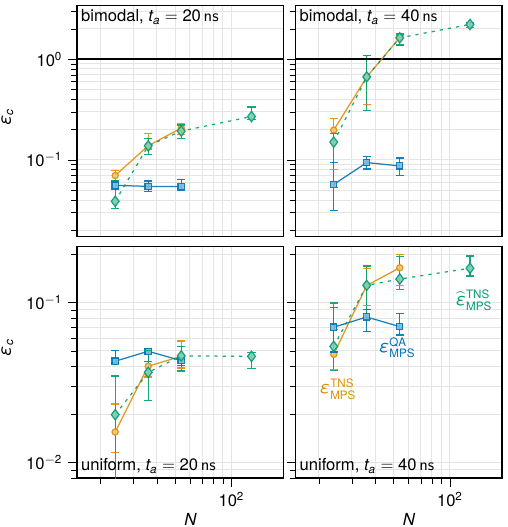}
\caption{{\bf Scaling of error with system size in cubic dimer inputs.}  Median correlation error $\epsilon_c$ is shown for BP-TNS ($\circ$) and QA ({\tiny $\square$}) against an MPS ground truth, i.e.~$\epsmpstns$ and $\epsmpsqa$ respectively.  Also shown is the indirect estimator $\hat{\epsilon}{}_\text{MPS}^\text{TNS}$ ($\diamond$), which extends the reach beyond sizes simulatable via MPS.  BP-TNS data have $\chi=8$ and $\lmax=7$.  From $L=3$ to $L=4$ ($N=54$ to $N=128$), $\hat{\epsilon}{}_\text{MPS}^\text{TNS}$ increases for bimodal spin glasses and is flat (within error bars) for uniform spin glasses.}\label{fig:4}
\end{figure}

Fig.~\ref{fig:4} shows scaling of different errors with size, including the QPU-estimated ensemble-median error $\hat{\epsilon}{}_\text{MPS}^\text{TNS}$. As expected, $\hat{\epsilon}{}_\text{MPS}^\text{TNS}$ agrees with  $\epsmpstns$ in most points, especially when quantum and classical errors are not close. In all cases, $\hat{\epsilon}{}_\text{MPS}^\text{TNS}$ increases with $N$, except for the uniform case at $t_a=\SI{20}{ns}$ for which the last two points are flat. There is no evidence of $\hat{\epsilon}{}_\text{MPS}^\text{TNS}$ decreasing from $L=3$ to $L=4$ (between the last two points), within error bars, over the ensembles tested. This disagrees with the predictions of Ref.~\cite{tindall_dynamics_2025}.  For bimodal inputs, we see evidence that the error continues to increase up to $L=4$ and possibly beyond, despite the disappearance of the shortest (three-site) loops that exist for $L=3$.  

\begin{figure*}
\includegraphics{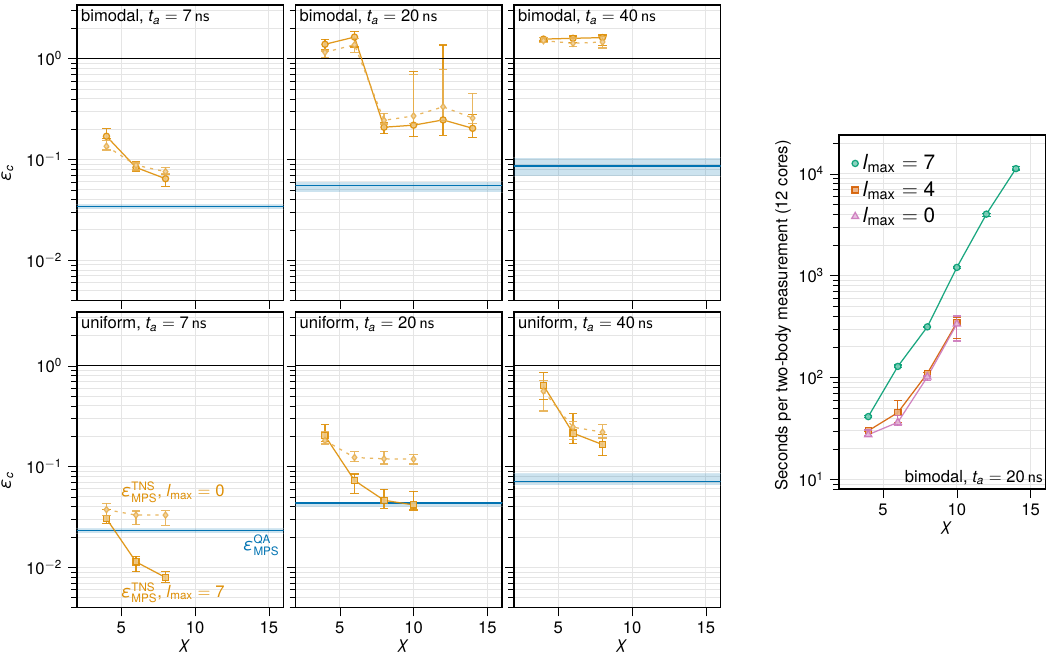}
\caption{{\bf Scaling of BP-TNS error with bond dimension in cubic dimer inputs.}  Left: Median $\epsmpstns$ is plotted against bond dimension $\chi$ for $L=3$ inputs.  Ensemble median and 95\% confidence intervals are indicated by markers and error bars for BP-TNS, and by lines and shaded areas for QA.  Right: Scaling of CPU time per two-body measurement as a function of bond dimension $\chi$.}\label{fig:5}
\end{figure*}

\section{Discussion}

In this work we have demonstrated utility of a quantum processor in benchmarking classical algorithms in a regime where generating a classical ground truth of equal quality to the QPU would be impractical.  In doing so, we have shed new light on the scaling of error in BP-TNS.  We see the expected degradation of performance of BP-TNS for longer simulated quenches, which is pronounced in bimodal spin glasses.  We also found that loop corrections, which are very effective in models with short correlation length and without short loops, become ineffective and even counterproductive with increasing loop contributions, and thus should not be assumed to generically improve estimators. Furthermore, it appears that increasing the BP-TNS bond dimension also offers diminishing returns (Fig.~\ref{fig:5}, left), underscoring the limitation of belief propagation algorithms in approximating contractions of tensor networks with short loops.  Although measurement time should asymptotically scale as $\chi^7$ for cubic dimer lattices, it appears to scale exponentially over the range studied (Fig.~\ref{fig:5}, right).

As in Ref.~\cite{tindall_dynamics_2025}, we have only investigated fidelity of spin-spin correlations; the distribution quality of full samples that can be produced using BP-TNS has only been explored very recently~\cite{rudolph_simulating_2025} in the context of open-boundary 2D quantum circuits.

While this demonstration of quantum utility is itself partly motivated by the question of the QPU's own accuracy, evaluating novel classical simulation methods remains an important task.  Moreover, simulating the quantum quench dynamics of Ising spin glasses has proven to be a useful computational challenge for establishing the limits of classical simulability.

\bibliography{main}

\section*{Acknowledgments}  The authors are grateful to Joseph Tindall for helpful discussions.  Work at UBC was supported by Natural Sciences and Engineering Research Council of Canada (NSERC) Alliance Quantum Program (Grant ALLRP-578555), CIFAR and the Canada First Research Excellence Fund, Quantum Materials and Future Technologies Program. This research used computational resources and services provided by Advanced Research Computing at the University of British Columbia. It also used resources of the Oak Ridge Leadership Computing Facility, which is a DOE Office of Science User Facility supported under Contract DE-AC05-00OR22725.

\clearpage
\begin{center}
\textbf{\large Supplementary Materials:\\ \mytitle}
\end{center}

\setcounter{equation}{0}
\setcounter{figure}{0}
\renewcommand{\figurename}{FIG.}
\renewcommand{\thefigure}{S\arabic{figure}}
\renewcommand{\theequation}{S\arabic{equation}}
\renewcommand{\theHfigure}{S\arabic{figure}}
\renewcommand{\thefootnote}{{\Roman{footnote}}}
\makeatother

\section{Spin-glass ensembles}

Here we study two types of Ising spin-glass ensembles based on simple cubic and diamond spin glasses, using the same family of inputs as Ref.~\cite{king_beyondclassical_2025}.  The coupling value $J_{ij}$ for adjacent simple cubic or diamond lattice sites $i$ and $j$ is drawn either from the bimodal uniform distribution over $\{-1,1\}$ as in the ``bimodal'' model, or the uniform distribution over the values $[-1,1]$, discretized to a step size of $\tfrac 1{256}$ and excepting $0$, in the ``uniform'' model.  In the cubic case, lattice sites are replaced with two-qubit dimers to allow embedding into the respective Pegasus and Zephyr qubit graphs of Advantage\texttrademark\ and Advantage2\texttrademark\ QPUs~\cite{king_quantum_2023,king_beyondclassical_2025}, and the two sites in a dimer are coupled with $J_{ij}=-2$.

\section{Quantum annealing methods}\label{sec:qa}

The QA experiments in this work were run on the same Advantage2\texttrademark\  prototype used in Ref.~\cite{king_beyondclassical_2025} (ADV2).  As in Ref.~\cite{king_beyondclassical_2025}, this system was used to simulate a model quench based on the annealing schedule of an older-generation Advantage\texttrademark\ QPU.  Doing this involves calibrating the appropriate experimental time $\hat t_a$ to match a given model $t_a$ (of $7$, $20$, or $\SI{40}{ns}$ in this work); we do so as follows.

For an ensemble of 20 $4{\times}6$ square-lattice uniform spin glasses, we perform MPS simulations at a range of model $t_a$.  For a range of energy scales $\alpha$, we multiply the classical spin-glass Hamiltonian $\mathcal H_I$ by $\alpha$ and quench at $\SI{5}{ns}$, the fastest allowed by the QPU.  For each $\alpha$, we find the best-fit model $t_a$ based on ensemble-median $\langle q^2\rangle$, and denote this experimental time $\hat t_a(\alpha)$: the model time best approximated by a $\SI{5}{ns}$ anneal at energy scale $\alpha$.  Then given a model time $t_a$ and an energy scale $\alpha$, we use an experimental annealing time of $\hat t_a(\alpha)*t_a/(\SI{5}{ns})$.  Using smaller $\alpha$ results in larger experimental annealing time, which can bring with it effects of decoherence and analog error.  Using large $\alpha$ pushes the critical point earlier in the anneal, where deviations between qubit and ideal spin-1/2 behaviour are greater.  We therefore choose a constant $\alpha$ between $0.4$ and $1.0$ depending on $t_a$ and the variance of the random couplings (i.e., uniform or bimodal), independent of system size and geometry.

For each spin-glass realization and parametrization ($\alpha$, $t_a$), we repeat $1000$-shot QPU calls while adjusting a flux-offset shim to keep qubits balanced at zero magnetization~\cite{chern_tutorial_2023}.  Each QPU call simulates multiple embeddings of the spin-glass realization in parallel, for example 3 embeddings for cubic dimer lattices with $L=4$, yielding $3{,}000$ logical shots per call.  In each QPU call and for each embedding, the spin-glass realization is embedded subject to a random graph automorphism, to suppress errors arising from device variation.

\section{Software methods}

Our BP-TNS experiments were performed using the TensorNetworkQuantumSimulator package~\cite{tensornetworkquantumsimulator} in conjunction with iTensorNetworks.jl~\cite{itensornetworks} and iTensor~\cite{fishman_itensor_2022}, and generally repeat the methodology of Ref.~\cite{tindall_dynamics_2025}.  Simulations were run from $s=0$ to $s=0.6$, which has been established as being close enough to $s=1.0$ as to not contribute significant error in these specific experiments (\cite{king_beyondclassical_2025} SM Fig.~S23).  We used a Trotterized circuit with a time step of $\SI{0.01}{ns}$, in which each dimer is evolved as a single degree of freedom rather than two distinct spins.  We found it necessary to update the BP message cache upon each gate application, rather than each layer as in~\cite{tindall_dynamics_2025}, to avoid numerical instabilities late in the anneal, particularly for bimodal inputs.

The dominant computational cost for BP-TNS estimates is not time evolution, but rather the estimation of correlation operators from the final wave function.  To reduce computational resource requirements in cubic dimer inputs, we computed only spin-spin correlations involving the first spin of each dimer, and expanded these correlations to a full set of spin-spin correlations by assuming that two spins in any dimer are perfectly correlated.  This is a good approximation in uniform spin glasses, but not necessarily true in bimodal spin glasses, where ground-truth dimer correlations as low as $0.7$ were found.  However, the mean dimer correlation at $\SI{20}{ns}$ for $L=3$ is $0.991$, so we do not expect the deviation to significantly effect estimates of correlation error; this approximation reduces computational cost by a factor 4.

MPS experiments repeat the methodology of Ref.~\cite{king_beyondclassical_2025} and were run on the Frontier supercomputer at the Oak Ridge Leadership Computing Facility.

\end{document}